\documentstyle[amssymb,aps]{revtex}

\newtheorem{theorem}{Theorem}
\newtheorem{acknowledgement}[theorem]{Acknowledgement}

\begin{document}
\title{{\bf Theory of self phase-locked optical parametric oscillators}}
\author{J.-J. Zondy and A.\ Douillet}
\address{Laboratoire Primaire du Temps et des Fr\'{e}quences\\
Bureau National de M\'{e}trologie / Observatoire de Paris\\
61, avenue de l'Observatoire, F-75014 Paris, France.}
\author{A. Tallet, E. Ressayre and M. Le Berre}
\address{Laboratoire de Photophysique Mol\'{e}culaire\\
Batiment 210, Universit\'{e} de Paris Sud, 91405, Orsay Cedex, France.}
\date{\today}
\maketitle

\begin{abstract}
The plane-wave dynamics of $3\omega \rightarrow 2\omega ,\omega $
subharmonic optical parametric oscillators containing a second harmonic
generator of the idler-wave $\omega $ is analyzed analytically by using the
meanfield approximation and numerically by taking into account the field
propagation inside the media. The resonant $\chi ^{(2)}(-3\omega ;2\omega
,\omega )$ : $\chi ^{(2)}(-2\omega ;\omega ,\omega )$ cascaded second-order
nonlinearities induce a mutual injection-locking of the signal and idler
waves that leads to coherent self phase-locking of the pump and subharmonic
waves, freezing the phase diffusion noise. In case of signal-and-idler
resonant devices, largely detuned sub-threshold states occur due to a
subcritical bifurcation, broadening out the self-locking frequency range to
a few cavity linewidths.

\end{abstract}


\bigskip

\section{\bf Introduction}

The synthesis of phase-coherent (sub)harmonic optical radiation ($\omega
,2\omega ,...,N\omega $) is useful in high precision optical measurements,
such as optical frequency metrology \cite{Hansch/Telle}, \cite{Wong}. In
frequency metrology, a phase-coherent optical {\em by-N} divider allows to
reduce the absolute measurement of an optical frequency $N\omega $ (e.g,
couting cycles in the hundreds of THz range) to the measurement of a smaller
frequency interval $\Delta =\omega $ by use of a femtosecond laser comb
generator whose radio-frequency intermode spacing is phase-locked to a
primary microwave clock \cite{Hall/Hansch}. The generation of a comb of
phase-locked harmonic radiation can also be the starting basis for the
generation of attosecond pulse train by Fourier synthesis \cite{HanschOC},
\cite{Mukai},\cite{Kobay1}. Indeed, the superposition (Fourier synthesis\cite
{Siegman},\cite{Hyodo}) of a comb of {\em N} equal amplitude optical
harmonic fields with a controllable relative phase relationship can lead to
a temporal train of ultra-short pulses with a sub-optical cycle duration ($%
\tau \thicksim 2\pi /N\omega $) and a repetition rate equal to the
fundamental harmonic $\omega $. Phase-locked optical harmonic generators are
often based on many lasers linked via nonlinear upconversion parametric
processes \cite{Hansch/Telle}, \cite{Pfister},\cite{Touahri}, \cite{Nee1},
\cite{Bernard}. Phase coherence among the harmonic waves is usually achieved
by use of complex electronic phase-locking loops. The high conversion
efficiencies of optical parametric oscillators (OPOs) combined with the
engineering flexibility offered by periodically poled (PP) nonlinear
materials may allow a compact implementation of such optical subharmonic
generators or attosecond pulse generator.

The study of OPOs as by-$N$ dividers ($N=2,3,4$) of a pump frequency $\nu
_{p}$ is motivated by their capacity to perform the phase-coherent division
of a pump photon into two highly phase-correlated subharmonic photons. In
precision measurement setups, subharmonic generation leads to a subsequent
phase noise reduction ($\div N^{2}$) compared to harmonic generation ($%
\times N^{2}$). Graham and Haken have first demonstrated that the phase sum $%
\varphi _{1}+\varphi _{2}$ of the idler and signal waves from an OPO follows
adiabatically the phase noise $\varphi _{p}$ of the pump laser, while their
phase difference $\varphi _{2}-\varphi _{1}$ undergoes a phase diffusion
process stemming from the interaction of both modes with the vacuum
fluctuations\cite{Graham}. To act as pertinent phase-coherent dividers, the
idler and signal waves have thus to be phase-locked, for instance by use of
an electronic servo control that forces $\varphi _{2}-\varphi _{1}$ to copy
the phase of an external RF oscillator referenced to a primary clock. Such
an electronic servo technique has been implemented to develop phase-coherent
divide-by-2 (2:1)\cite{Lee}, and divide-by-3 (3:1) OPOs for high resolution
optical metrology \cite{Slyusarev}, \cite{Douillet} with corresponding
residual phase difference variance well below $1$ rad$^{2}$.

Electronic phase-locking loops are however subject to limited response time
bandwidth and require fast OPO cavity length actuators. As an alternative,
all-optical self phase-locking (SPL) methods are currently being
investigated \cite{Mason},\cite{Boller} to by-pass the bandwidth limitation
of electronic servos and simplify the implementation of phase-locked
dividers. These methods are based on the self injection-locking of the
subharmonic waves, similarly to the injection locking process of a slave
laser oscillator by a master laser which possesses a better spectral purity
and frequency stability \cite{Siegman},\cite{Nabors}. In injection-locked
systems, the phase of the slave oscillator, $\phi (t)=\omega t+\varphi (t)$,
is a perfect copy of the phase of the injecting master oscillator. Phase
locking can occur only when the two frequencies are close enough, e.g.
within a certain {\em locking range} $\Delta \omega _{lock}=\left| \omega
_{s}-\omega _{m}\right| $ whose extent is proportional to the squared ratio $%
\sqrt{P_{m}/P_{s}}$ of the injecting master power to the slave power. It is
thus expected that the self-locking range in an OPO divider should depend on
whether the OPO is configured as a doubly-resonant (DRO), a triply-resonant
(TRO) or a pump-enhanced singly resonant oscillator (PRSRO). Strong self
injection-locking regime occurs only when the signal and idler waves are
simultaneously resonant.

For divide-by-2 OPOs based on a type-I nonlinear process (the signal and
idler are identically polarized), self-injection locking occurs naturally
when the OPO is tuned close enough to the frequency-degeneracy\cite
{NaborsOPO}. Recently a type-II self phase-locked 2:1 DRO was demonstrated,
in which a small mixing of the two orthogonally polarized signal and idler
waves, performed by an intracavity wave plate, induces the injection locking
\cite{Mason}. In both cases the evidence of self phase locking was provided
by the high level of phase-coherence between the frequency degenerate signal
and idler waves. Following the experiment in ref \cite{Mason}, the theory of
{\em linearly} coupled type-II 2:1 SPL-DROs has been reported\cite
{Fabre-Wong}. In the case of 3:1 OPO dividers, obviously such a linear
coupling cannot be implemented, and one must use a nonlinear coupling via $%
\chi ^{(2)}(-3\omega ;2\omega ,\omega ):\chi ^{(2)}(-2\omega ;\omega ,\omega
)$ cascading processes (for instance by introducing inside the OPO cavity a
second nonlinear material phase-matched for the SHG of the idler wave). The
first implementation of such a self \ phase-locked (SPL) OPO was recently
reported, and used a dual-grating periodically poled lithium niobate chip
carrying the OPO and SHG sections in a pump-resonant idler-resonant
configuration (PRSRO) \cite{Boller}. An extremely small locking range of $%
\thicksim $500 kHz, corresponding to a fraction of the idler cavity
linewidth, was reported due to the weak (non resonant) doubled idler power.
A different situation should occur when both the signal and idler waves are
resonant, as in a DRO or a TRO. Because of the enhanced energy flow exchange
between the $2\omega $ and $\omega $ modes, the dynamics of such a
self-phase locked OPO (SPL-OPO) is expected to differ significantly from
that of a conventional OPO. Strictly speaking the 3:1 SPL-DRO or TRO can be
regarded as a degenerate sub/second harmonic generator ($2\omega
\leftrightarrows \omega +\omega $) internally pumped by the signal wave of a
non-degenerate OPO ($3\omega \leftrightarrows 2\omega +\omega $). While
theories of OPO devices containing additional up or down conversion of the
signal and idler waves have been investigated in the past with the aim of
generating new frequencies \cite{Tang},\cite{Cheung},\cite{Moore},\cite
{Dikmelik}, or revealing the quantum noise signature of competing $\chi
^{(2)}$ nonlinearities \cite{Schiller-casc},\cite{Marte} such a
subharmonically resonant configuration of competing nonlinearities in OPOs
has never been theoretically investigated in detail. A 3:1 DRO-SHG rough
rate equation analysis based on photon flux conservation, but neglecting the
role of cavity detunings and field phase coherence, was previously given by
Zhang {\it et al} \cite{Kobayashi} with the conclusion that self-phase
locking should manifest through an imbalance of the signal and idler
intensities at exact 3:1 division. Actually our in-depth analysis shows that
such an imbalance holds only for zero detunings: When the nonlinear phase
shift due to cascading is compensated by a non-zero optimal cavity detuning,
one can obtain the same output intensities as in a conventional DRO.

In this paper, we provide a detailed plane-wave description of the dynamics
of divide-by-3 SPL-OPOs that encompasses all the three cavity configurations
(only the main results are summarized in the conclusion for the PRSRO, which
will be detailed elsewhere). The dynamics of the nonlinearly coupled OPO is
shown to differ substancially from the linear coupling case of a type-II 2:1
SPL-DRO \cite{Fabre-Wong}. In particular, the nonlinear coupling gives rise
to a subcrital bifurcation for any non-zero cavity detuning while
conventional OPOs undergo subcritical bifurcations only for the case of
largely pump-detuned TROs \cite{Lugiato}, \cite{Debuisschert}. As a
consequence of the nonlinear phase shifts due to the cascading processes,
SPL-DROs or TROs will be shown to display large-intensity stable stationary
states that would correspond to non-lasing detuning domains in conventional
OPOs, extending thus the phase-locking detuning range up to a few cavity
linewidths.

This paper is organized as follows. In section II we describe the basic
plane-wave ring cavity model leading to the meanfield solutions that will be
compared with the full propagation solutions. In section III, we first treat
in detail the DRO configuration case. We then extend the theory to the case
of a TRO in section IV. Main results on the PRSRO configurations will be
also summarized in the conclusion. Section V discusses the practical
implementation of such SPL-OPOs on the basis of the theoretical findings and
section VI concludes with some prospective studies aroused by this first
theoretical approach.

\noindent \qquad

\section{\bf Basic model and exact numerical solutions}

The OPO-SHG device described in the paper is schematically sketched in {\bf %
Fig.1}. We consider a ring cavity containing the cascaded OPO and SHG
nonlinear media, each of length $L_{1\text{ }}$and $L_{2}$ respectively and
assume that the pump, signal and idler waves satisfy already the 3:2:1
frequency ratios. Though the drawing depicts the case of a dual-grating PP
material, phase-matched for an ($eee$) interaction, the model is valid also
for any separate birefringent material sections, or a single material
phase-matched for both interactions. In the case of birefringent
phase-matching however, the OPO should be of type-II kind (e.g. $%
e\rightarrow o+e$ or $o\rightarrow e+o$) and the SHG of type-I kind (resp. $%
e+e\rightarrow o$ or $o+o\rightarrow e$), so that the system is described by
only 3 field variables.

\subsection{\bf Plane-wave propagation equations}

We denote by $z=Z/L_{1}$ and $z^{\prime }=Z/L_{2}$ the normalized
propagation distances within each crystal, such that $0\leq z\leq 1$ between
points ''0'' and ''1'' and $0\leq z^{\prime }\leq 1$ between ''1'' and
''2''. Let $E_{j}$ (in m/V) be the slowly varying complex field amplitudes
where the subscripts $j=p,2,1$ stand respectively for pump, signal and idler

\begin{equation}
\epsilon _{j}(Z,t)=\frac{1}{2}E_{j}(Z,t)\exp \left[ i(\omega _{j}t-k_{j}Z)%
\right] +c.c.  \eqnum{1}
\end{equation}
and let $N_{j}$ be the complex field variable, such that $\left|
N_{j}\right| ^{2}$ is the number of photons in mode $j$ at plane $z$ inside
the ring cavity, by

\begin{equation}
N_{j}=\sqrt{\frac{\epsilon _{0}cV}{2\hslash \omega _{j}}}E_{j}  \eqnum{2}
\end{equation}
where $V$ is the average volume occupied by the modes inside the resonator.
Then, the reduced field amplitudes $A_{j}$ are introduced by scaling $N_{j}$
with the small signal gain coefficient $g_{1}L_{1}$ of the OPO crystal

\begin{equation}
A_{j}=g_{1}L_{1}N_{j}\   \eqnum{3}
\end{equation}

with

\begin{equation}
g_{1}=\frac{d_{OPO}}{c}\sqrt{\frac{2\hslash \omega _{1}\omega _{2}\omega _{p}%
}{\epsilon _{0}Vn_{1}n_{2}n_{p}}}  \eqnum{4}
\end{equation}

In Eq.(4), $d_{OPO}={\bf e}_{p}\cdot \lbrack \chi ^{(2)}(-3\omega ;2\omega
,\omega )/2]{\bf :e}_{2}\cdot {\bf e}_{1}$ (in m/V) is the effective
nonlinear coefficient for the OPO interaction and $n_{j}$ are the refractive
indices at frequency $\omega _{j}$. The reduced Maxwell equations for these
field amplitudes $A_{j}$ (z,t), when propagating through the ($3\omega
\rightarrow 2\omega ,\omega )$ crystal without any diffraction effect, are

\begin{eqnarray}
\frac{dA_{p}}{dz} &=&iA_{1}A_{2}  \eqnum{5a} \\
\frac{dA_{2}}{dz} &=&iA_{p}A_{1}^{\ast }  \eqnum{5b} \\
\frac{dA_{1}}{dz} &=&iA_{p}A_{2}^{\ast }  \eqnum{5c}
\end{eqnarray}
using the standard slowly varying amplidude approximation and the usual
change of variables $z\longrightarrow z,t\longrightarrow t-\overline{n}z/c,$
where $\overline{n}$ is the mean linear refractive index. In Eqs. 5 perfect
phase-matching has been assumed, $\Delta k_{OPO}=k_{p}-k_{2}-k_{1}=0$. The
propagation equations in the SHG crystal, allowing for a non-vanishing
wavevector mismatch $\Delta k_{SHG}=k_{2}-2k_{1}$, are

\begin{eqnarray}
\frac{dA_{p}}{dz^{\prime }} &=&0  \eqnum{6a} \\
\frac{dA_{2}}{dz^{\prime }} &=&iSA_{1}^{2}\exp (+i2\xi z^{\prime })
\eqnum{6b} \\
\frac{dA_{1}}{dz^{\prime }} &=&iSA_{2}A_{1}^{\ast }\exp (-i2\xi z^{\prime })
\eqnum{6c}
\end{eqnarray}

with initial conditions $A_{j}(z^{^{\prime }}=0)=A_{j}(z=1).$

The two cascaded crystals are usually phase-matched by adjusting their
temperature or angle. But, exact zero mismatch for both interactions may be
difficult to achieve, so that a phase-mismatch parameter $\xi =\Delta
k_{SHG}L_{2}/2$ is introduced. The parameter $S$ in eqs. (6) is the ratio of
the SHG to OPO small signal gains

\begin{equation}
S=\frac{g_{2}L_{2}}{g_{1}L_{1}}  \eqnum{7}
\end{equation}

where the small signal gain SHG coefficient (in m$^{-1}$) is

\begin{equation}
g_{2}=\frac{d_{SHG}}{c}\sqrt{\frac{2\hslash \omega _{1}^{2}\omega _{2}}{
\epsilon _{0}Vn_{1}^{2}n_{2}}}  \eqnum{8}
\end{equation}

For a type-I (eee) PP crystal the effective nonlinear coefficients satisfy $%
d_{OPO}\approx d_{SHG}$, so that with $n_{1}\approx n_{2}\approx n_{p}$ and $%
\omega _{p}=3\omega $, $\omega _{2}=2\omega $, $\omega _{1}=\omega ,$ the
relation (7) can be approximated by $S\approx (L_{2}/L_{1})/\sqrt{3}$. For $%
L_{2}/L_{1}=1/3$ for instance, one has $S\approx 0.2$.\bigskip

The solutions at the SHG crystal exit can be limited to few terms of the
Mac-Laurin field expansion serie at the cell entrance, because of the
smallness of the parametric gain coefficients. We shall refer to ML1
approximation, keeping only the lowest-order perturbative terms, i.e.
quadratic terms in powers of the field amplitudes. At the exit of the SHG
crystal, the ML1 axpproximation provides the solutions for the signal and
idler amplitudes as functions of their value at the entrance of the OPO
crystal (point ''0'' in Fig.1)

\begin{eqnarray}
A_{p}(t,L_{1}+L_{2}) &=&A_{p}(t,0)+iA_{1}(t,0)A_{2}(t,0)  \eqnum{9a} \\
A_{2}(t,L_{1}+L_{2}) &=&A_{2}(t,0)+iA_{p}(t,0)A_{1}^{\ast }(t,0)+i\chi
^{\ast }A_{1}^{2}(t,0)  \eqnum{9b} \\
A_{1}(t,L_{1}+L_{2}) &=&A_{1}(t,0)+iA_{p}(t,0)A_{2}^{\ast }(t,0)+i\chi
A_{2}(t,0)A_{1}^{\ast }(t,0)  \eqnum{9c}
\end{eqnarray}
with the nonlinear coupling constant $\chi $

\begin{equation}
\chi =S\exp (-i\xi )(\sin \xi /\xi )  \eqnum{10}
\end{equation}

\subsection{\bf Boundary conditions}

The reduced Maxwell equations have to be completed by the boundary
conditions for the three waves at the entrance of the OPO crystal to derive
the cavity equations. The field $A_{j}(t+\tau ,0)$ at location ''0'' and at
a time $t+\tau $ , where $\tau =\Lambda /c$ is the cavity roundtrip time ($%
\Lambda $ is the total cavity optical path), is the field $%
A_{j}(t,L_{1}+L_{2})$ at point ''2''of Fig.1, which propagates freely after
bounces at the totally reflecting ($R=1$) and output coupling ($r_{j},t_{j}$%
) mirrors and eventually summed with an input field, as it is the case for
the pump field. We define by $r_{j\text{ }}(t_{j})$ the overall mirror
amplitude reflectivities (transmissivities), such that $%
r_{j}^{2}+t_{j}^{2}=1 $ and define the amplitude loss coefficients $\kappa
_{j}=1-r_{j}$ .

Then, the boundary conditions take then the form

\begin{eqnarray}
A_{p}(t+\tau ,0) &=&r_{p}\exp (i\Delta _{p})A_{p}(t,L_{1}+L_{2})+A_{in}
\eqnum{11a} \\
A_{2}(t+\tau ,0) &=&r_{2}\exp (i\Delta _{2})A_{2}(t,L_{1}+L_{2})  \eqnum{11b}
\\
A_{1}(t+\tau ,0) &=&r_{1}\exp (i\Delta _{1})A_{1}(t,L_{1}+L_{2})  \eqnum{11c}
\end{eqnarray}

where $A_{in}$ denotes the input pump amplitude inside the cavity, $%
A_{in}=t_{p}A_{in}^{ext}$ . Each $\Delta _{j}$ is the usual cavity detuning
between the frequency of the waves and the corresponding cold cavity
frequency, scaled to the HWHM cavity resonance width ( Only nearly resonant
waves will be considered, $\Delta _{j}<<2\pi $).

The set of equations (9) and (11) forms a mapping of the field amplitudes at
location ''0'', from which meanfield equations may be derived.

The {\em meanfield model} was originally derived from a similar set of
equations for a two-level cell ring cavity device when the atomic dephasing
time is much greater than the round-trip time $\tau $, but much smaller than
the photon lifetime $\tau /\kappa $ \cite{Meanf1}. For an OPO, the situation
is somewhat different because the response of the crystal is assumed to be
instantaneous (see Eqs. 5). Nevertheless, meanfield equations can be also
derived from the ML1 equations (9) and the boundary conditions (11) if the
field amplitudes are slowly varying during a round-trip time$,$ i.e. $\tau
(dA_{j}/dt)<<A_{j}(t)$ \cite{Meanf2},\cite{Meanf3}.

In the conventional TRO \ case ($\left| \chi \right| =0$ , $\Delta
_{j}<<1,\kappa _{j}<<1),$ nontrivial homogeneous stationary solutions exist
for the idler and the signal only if the following relation holds\cite
{Lugiato},\cite{Debuisschert}
\begin{equation}
\stackrel{\_}{\Delta }=\Delta _{1}/\kappa _{1}=\Delta _{2}/\kappa _{2}.
\eqnum{12a}
\end{equation}

and the intracavity pump intensity $I_{p}=\left| A_{p}\right| ^{2}$ is
clamped to the threshold input intensity,

\begin{equation}
I_{th}=\kappa _{1}\kappa _{2}+\Delta _{1}\Delta _{2}  \eqnum{12b}
\end{equation}
whatever the input intensity $\left| A_{in}\right| ^{2}$ might be above the
threshold. In the case of the DRO, the relation (12a) still holds. But
solutions for the signal and idler intensities can be determined only if the
expansion (9) is continued up to the cubic terms, leading to

.
\begin{eqnarray}
I_{1} &=&2\kappa _{2}\left[ \sqrt{1+(1+\stackrel{\_}{\Delta }
^{2})(I_{p}/I_{th}-1)}-1\right]  \eqnum{12c} \\
I_{1}/I_{2} &=&\Delta _{2}/\Delta _{1}=\kappa _{2}/\kappa _{1}  \eqnum{12d}
\\
\cot (\varphi _{p}-\varphi _{1}-\varphi _{2}) &=&-\stackrel{\_}{\Delta }
\eqnum{12e}
\end{eqnarray}

The bifurcation is supercritical for any detuning . Besides, the phase
difference $\varphi _{1}-\varphi _{2}$ is an undetermined quantity while the
sum phase $\varphi _{1}+\varphi _{2}$ depends on the pump laser phase \cite
{Fabre-Wong}. Actually, this classical phase indetermination is compatible
with the result of the quantum fluctuation theory of parametric oscillators
\cite{Graham}.

\section{\bf SPL-DRO case}

\subsection{\bf Stationary solutions}

In the case of a DRO, only the signal and idler waves resonate with the
cavity while the pump wave is a travelling wave ($t_{p}=1$, $r_{p}=0$).
Assuming $\kappa _{1,2}\ll 1,$ $\Delta _{1,2}\ll 1,$ Eqs.(9-11) provide the
stationary solutions, $A_{j}(t+\tau ,0)=A_{j}(t,0)\equiv A_{j},$which are
also those of the meanfield model,

\begin{eqnarray}
A_{p} &=&A_{in}  \eqnum{13a} \\
(\kappa _{2}-i\Delta _{2})A_{2} &=&iA_{p}A_{1}^{*}+i\chi ^{*}A_{1}^{2}
\eqnum{13b} \\
(\kappa _{1}-i\Delta _{1})A_{1} &=&iA_{p}A_{2}^{*}+i\chi A_{2}A_{1}^{*}
\eqnum{13c}
\end{eqnarray}

Eqs (13) can be conveniently solved by setting $A_{j}=\alpha _{j}\exp
(i\varphi _{j})$, giving rise to

\begin{eqnarray}
(\kappa _{2}-i\Delta _{2})\alpha _{2} &=&i\alpha _{p}\alpha _{1}\exp (i\mu
)+i\left| \chi \right| \alpha _{1}^{2}\exp (-i\eta )  \eqnum{14a} \\
(\kappa _{1}-i\Delta _{1})\alpha _{1} &=&i\alpha _{p}\alpha _{2}\exp (i\mu
)+i\left| \chi \right| \alpha _{1}\alpha _{2}\exp (+i\eta )  \eqnum{14b} \\
\mu &=&\varphi _{p}-\varphi _{2}-\varphi _{1}\;  \eqnum{14c} \\
\eta &=&\varphi _{2}-2\varphi _{1}-\xi  \eqnum{14d}
\end{eqnarray}

The non-trivial solutions ($A_{j}$ $\neq 0)$ can be easily handled by
setting $X=(\left| \chi \right| \alpha _{1})^{2}$ and $I_{p}=\alpha
_{p}^{2}. $ Then the scaled idler intensity is found to be solution of
\begin{equation}
X^{2}-2(I_{p}-\kappa _{1}\kappa _{2}+\Delta _{1}\Delta _{2})X+\left| (\kappa
_{1}-i\Delta _{1})(\kappa _{2}+i\Delta _{2})-I_{p}\right| ^{2}=0  \eqnum{15}
\end{equation}

Let us note that the coupling terms in Eqs.(14) allow to obtain the
non-trivial solutions at the ML1 approximation order, unlike for the
conventional DRO.

Three cases have to be distinguished in order to solve Eq$.$ (15), $\Delta
_{1,2}=0,$ $\Delta _{1}\Delta _{2}\neq 0$ (but with $\Delta _{1}\Delta
_{2}>0 $)$,$ and $\Delta _{1,2}=0$, $\Delta _{2,1}\neq 0.$

For $\Delta _{1}\Delta _{2}\neq 0$, there are two solutions

\begin{equation}
X_{\pm }=(\left| \chi \right| \alpha _{1})_{\pm }^{2}=I_{p}-\kappa
_{1}\kappa _{2}+\Delta _{1}\Delta _{2}\pm 2\sqrt{\Delta _{1}\Delta _{2}\left[
I_{p}-I_{0}\right] }  \eqnum{16}
\end{equation}

for any input $I_{p}\geq I_{0},$ where $I_{0}$ is the input intensity at the
saddle-node bifurcation

\begin{equation}
I_{0}=\frac{1}{4}\left( \kappa _{1}\sqrt{\frac{\Delta _{2}}{\Delta _{1}}}
+\kappa _{2}\sqrt{\frac{\Delta _{1}}{\Delta _{2}}}\right) ^{2}  \eqnum{17}
\end{equation}

while the threshold intensity, determined by $dX_{_{-}}/dI_{p}$ $=0$, is

\begin{equation}
I_{th}=I_{0}+\Delta _{1}\Delta _{2}  \eqnum{18}
\end{equation}

which is minimum for $\Delta _{1}/\kappa _{1}=\Delta _{2}/\kappa _{2}$ and
equal to $\kappa _{1}\kappa _{2}+\Delta _{1}\Delta _{2},$ like for the
conventional DRO.

The signal intensity $I_{2}$ is related to $I_{1}$ via the relation
\begin{equation}
(\alpha _{1}/\alpha _{2})^{2}=\Delta _{2}/\Delta _{1}  \eqnum{19}
\end{equation}

deduced from Eqs.(14a-b), so that the two detunings must have the same sign.

The solutions $X_{_{+}}$ and $X_{_{-}}$ are represented by the lines a and
a', respectively, in {\bf Fig. 2} that displays a subcritical bifurcation:
Indeed, only the stationary solution $X_{_{+}}$ is stable ( solid line a )
and extends from the saddle-node intensity $I_{0}$ to the Hopf bifurcation
intensity $I_{H},$ above which the solution is periodic (dashed portion) .
The stationary solution $X_{_{-}}$ is marginally unstable (dashed line a'),
whatever the input intensity might be [See Appendix].

Let us notice that the conventional DRO detuning condition, $\stackrel{\_}{%
\Delta }=\Delta _{1}/\kappa _{1}=\Delta _{2}/\kappa _{2}$ does no longer
necessarily hold and hence the SPL-DRO can oscillate with a wider detuning
range. This result is similar to the result of a 2:1 degenerate SPL-DRO, in
the case of a linear coupling. However, this latter case does not display
any subcriticality: The linear coupling gives rise instead to two self
phase-locked states corresponding to two distinct thresholds, for a given
pumping rate\cite{Fabre-Wong}.

The phases $\mu $ and $\eta $ fulfill the relations

\begin{eqnarray}
\sin \mu &=&-\frac{1}{2\alpha _{p}}\left[ \kappa _{1}\sqrt{\frac{\Delta _{2}%
}{\Delta _{1}}}+\kappa _{2}\sqrt{\frac{\Delta _{1}}{\Delta _{2}}}\right] =-%
\sqrt{\frac{I_{0}}{I_{p}}}  \eqnum{20a} \\
\sin \eta &=&-\frac{1}{2\left| \chi \right| \alpha _{1}}\left[ \kappa _{1}%
\sqrt{\frac{\Delta _{2}}{\Delta _{1}}}-\kappa _{2}\sqrt{\frac{\Delta _{1}}{
\Delta _{2}}}\right]  \eqnum{20b} \\
\text{with }\alpha _{p}\cos \mu +\left| \chi \right| \alpha _{1}\cos \eta
&=&-\sqrt{\Delta _{1}\Delta _{2}}  \eqnum{20c}
\end{eqnarray}

Eqs.(20) completely determine the absolute phases of the subharmonic waves,
which are hence self-locked. These relations on the idler and signal phases,
while similar to those in ref \cite{Fabre-Wong}, will be shown to present
substantial differences. As a consequence of the self-locking, highly
phase-coherent subharmonic outputs are expected from the SPL-DRO.

\bigskip In the case $\Delta _{1,2}=0$, the bifurcation is supercritical
with
\begin{eqnarray}
X &=&I_{p}-I_{th}  \eqnum{21a} \\
I_{th} &=&\kappa _{1}\kappa _{2}  \eqnum{21b}
\end{eqnarray}
The real and imaginary parts of Eqs. (14a-b) give rise to

\begin{center}
\begin{eqnarray}
\sin \mu &=&-\frac{1}{2\alpha _{p}}\left[ \kappa _{1}\frac{\alpha _{1}}{
\alpha _{2}}+\kappa _{2}\frac{\alpha _{2}}{\alpha _{1}}\right]  \eqnum{22a}
\\
\sin \eta &=&-\frac{1}{2\sqrt{I_{p}-\kappa _{1}\kappa _{2}}}\left[ \kappa
_{1}\frac{\alpha _{1}}{\alpha _{2}}-\kappa _{2}\frac{\alpha _{2}}{\alpha _{1}%
}\right]  \eqnum{22b} \\
\alpha _{p}\cos \mu +\sqrt{I_{p}-\kappa _{1}\kappa _{2}}\cos \eta &=&0
\eqnum{22c}
\end{eqnarray}
\end{center}

Eq.(22c) implies that $\cos \mu $ and $\cos \eta $ are either of opposite
signs or simultaneously nil. The first case corresponds to

\begin{equation}
\frac{\alpha _{1}}{\alpha _{2}}=\frac{\kappa _{1}}{\kappa _{2}}=1\text{ with
}\sin \eta =0\text{ and }\sin \mu =-\frac{\kappa _{1,2}}{\alpha _{p}}
\eqnum{23}
\end{equation}

The case $\cos \mu =\cos \eta =0$ is compatible only with $\kappa _{1}\neq
\kappa _{2}$ , leading to
\begin{eqnarray}
\frac{\alpha _{1}}{\alpha _{2}} &=&\sqrt{\frac{\kappa _{2}}{\kappa _{1}}}%
\left( \sqrt{N}\mp \sqrt{N-1}\right)  \eqnum{24a} \\
\sin \eta &=&\pm 1\text{ and }\sin \mu =-1  \eqnum{24b}
\end{eqnarray}
where $N=I_{p}/\kappa _{1}\kappa _{2}$ is the pumping rate. Eq.(24a)
displays a dependence of $\frac{\alpha _{1}}{\alpha _{2}}$ on the pump
amplitude, unlike eq.(19) and consequently the solutions are not continuous
when $\Delta _{1,2}\rightarrow 0,$ in the general case $\kappa _{1}\neq
\kappa _{2}$ .

For $\Delta _{1,2}=0,$ $\Delta _{2,1}\neq 0,$ Eq. (15) has no real solution.
Actually the exact solutions will be shown to display a time-dependent
regime.

Finally let us note that the signal and idler intensities are inversely
proportional to $\left| \chi \right| ^{2}$ as displayed by Eq.(16). This
dependence is a consequence of the ML1 approximation, Eqs.(9). (The limit $%
\left| \chi \right| \rightarrow 0$ is irrelevant since the meanfield
equations (13) are misleading for the conventional DRO).

\subsection{\bf Numerical results}

The time evolution of the field amplitudes is obtained by solving the
propagation equations (5-6) with the boundary conditions (11) for a given
input amplitude and small initial signal and idler amplitudes. The solutions
are obtained by numerical integration of Eqs. (5-6) using of a fourth-order
Runge Kutta algorithm, until convergence is achieved. In most of the
calculations, the cavity loss coefficients are taken constant and equal, $%
\kappa _{1,2}=0.005,$ and the SHG phase mismatch $\xi =\Delta kL_{2}/2$ $=0$%
, unless otherwise stated. The input amplitude, the detunings and the
coupling constant are varied.

\begin{itemize}
\item  {\large Stationary solutions}
\end{itemize}

As predicted by the linear stability analysis of the meanfield solutions, a
single solution, corresponding to $X_{_{+}},$ occurs by numerical
integration for an input amplitude above the threshold given by Eq.(18),
where the trivial solution is unstable. In a general manner, the signal and
idler intensities satisfy the relations predicted by the mean-field, either
for non-zero detunings in Eq. (19) or vanishing detunings in Eqs. (23-24).
Nevertheless, for zero detunings, numerical intensities agree with the
meanfield intensities, only for small coupling parameters ($\left| \chi
\right| \leqslant 0.15$). As already pointed out, there is no continuity for
the intensities when $\Delta _{1,2}\rightarrow 0$ in the case $\kappa
_{1}\neq \kappa _{2}.$

The exact stationary amplitude $\left| \chi \right| \alpha _{1}$ and its
stability domain depend on the magnitude of $\left| \chi \right| $ as
displayed in Fig.\ 2 by line $b$ $\ $for $\left| \chi \right| =0.05$ and
line $c$ for $\left| \chi \right| =0.2.$ In this latter case, the exact
solution is close to the meanfield one, the stationary solution extends
almost from $I_{0}$ in the subthreshold domain, reachable only by backward
adiabatic decrementation of the pump intensity, to the Hopf bifurcation
threshold intensity $I_{H}^{num}$ , slightly larger than the meanfield value$%
.$ As $\left| \chi \right| $ decreases, the domain of stability is shortened
below threshold ($I_{0}^{num}>I_{0}$ ),but it is enlarged above threshold,
because the Hopf bifurcation threshold is shifted towards larger input
intensity, $I_{H}^{num}>I_{H}$ ( For $\kappa _{1,2}=\Delta _{1,2}=0.005,$
the meanfield predicts $I_{H}=2I_{th}$, while the numerical values are $%
I_{H}^{num}\simeq $ $2.6I_{th}$ for $\left| \chi \right| =0.2$ and $%
I_{H}^{num}>4I_{th}$ for $\left| \chi \right| =0.05)$.

Subcriticality is also evidenced in{\bf \ Fig.3}, when starting the
numerical integration from $\stackrel{\_}{\Delta }=0$ and performing an
adiabatic increase of $\stackrel{\_}{\Delta },$ for a fixed input pump
intensity $I_{p}=4\kappa _{1}\kappa _{2}.$ The different curves in solid
line correspond to different values of $\left| \chi \right| $. They are
symmetric for negative detunings, so that the tuning curve of the SPL-DRO
appears as a widened double-sided fringe, which reminds of resonantly
phase-modulated Fabry Perot devices (the same kind of fringe is obtained
from a resonator containing an electro-optic phase modulator driven by a RF
oscillator whose frequency is equal to the resonator free spectral range\cite
{OFCG}). In the case of the conventional DRO ($\left| \chi \right| =0$,
dashed curve), lasing begins at $\stackrel{\_}{\Delta }=0$ where the
intensity is maximum as seen in Eq.(12c) and stops for $\stackrel{\_}{\Delta
}=\sqrt{3;}$ indeed, the bifurcation is supercritical so that lasing may
occur only if the input pump intensity is larger than the threshold value $%
\kappa _{1}\kappa _{2}+\Delta _{1}\Delta _{2}.$ The case $\left| \chi
\right| =0.01$ (curve a), corresponding to nascent bistability, displays
lasing, approximately in the same range of detuning as in the $\left| \chi
\right| =0$ case, with the significant difference that there is self
phase-locking (see below on Fig.6). As $\left| \chi \right| $ increases
further, the bifurcation becomes subcritical, the saddle-node moves away
from the threshold, approaching the meanfield location, independently of $%
\stackrel{\_}{\Delta }$. The intensity reaches the same maximum $I_{\max }$
for any $0\leq \left| \chi \right| \leq 0.25,$ but at a detuning $\overline{%
\Delta }_{\max }($ $\left| \chi \right| ),$ for which the pump is entirely
depleted; then the intensity decreases to zero when increasing further the
detuning. The maximum $I_{\max }$ is equal to the maximum conventional DRO
intensity, which occurs at $\overline{\Delta }_{\max }=0$. In case of the
parameters used in Fig.3, Eq. (12c) gives rise to $\sqrt{I_{\max }}=0.1,$ in
agreement with the numerical result. The detuning $\overline{\Delta }_{\max
} $ becomes approximately proportional to $\left| \chi \right| $ for large
enough coupling strength. For $\left| \chi \right| \geq 0.25,$ the idler
intensity becomes time-dependent before $\overline{\Delta }_{\max }$ is
reached.

In summary, the cascaded SHG nonlinearity induces a $\left| \chi \right| -$
dependent phase that shifts the optimum detuning from $0$ to $\stackrel{\_}{%
\Delta }_{\max }.$

The temporal response of the system subject to stepwise-like detuning jumps
has been also studied in order to check the stability of the subthreshold
states against external perturbations that tend to modulate the detuning
parameter (via the cavity length for instance). {\bf Figure 4}, associated
with $\left| \chi \right| =0.2$, shows that the system recovers its steady
state operating point as long as the perturbation amplitude does not exceed $%
\pm 0.12\stackrel{\_}{\Delta }_{\max .}$ Note the longer decay time, which
is characteristic of a critical slowing-down phenomenon, on the positive
detuning step side that brings the DRO very close to the saddle-node
bifurcation.

We can define a self-locking domain in the ($\Delta _{1},\Delta _{2}$) plane
over which the DRO is self phase-locked with a well-defined phase
relationship. Contour plots of the numerical solutions for the idler $I_{1}$%
, the signal $I_{2}$, the total subharmonic intensity $I_{1}+I_{2}$ and the
phase difference $\eta $ are shown in {\bf Figs 5} for $\left| \chi \right| $
$=0.2$ and for a given pumping rate $N=I_{p}/\kappa _{1}\kappa _{2}=4$ . The
subharmonic intensities are scaled to the conventional zero-detuning DRO
values as deduced from Eqs.(12c), $I_{1,2}^{DRO}=2\kappa _{2,1}$. These
contour plots are obtained by adiabatical following of the stationary
solutions in order to reveal the 2D subthreshold domain. The granular small
regions adjacent to the $\Delta _{1}=0$ axis correspond to a time-dependent
regime. The $I_{1}$ and $I_{2}$ intensity distributions in Figs\ 5a-5b
display off-diagonal maxima that are larger than the conventional DRO signal
and idler outputs ( $I_{1,2}/I_{1,2}^{DRO}=1.6$). Note also that the exact
intensities $I_{1}$ and $I_{2}$ are not invariant with respect to the
product $\Delta _{1}\Delta _{2},$ differently from the predicted meanfield
intensities (16) but they satisfy the relation (19). The figure (5c) shows
that the maximum total intensity, slightly larger ($I_{1}+I_{2}=2.2$ ) than
the conventional DRO total intensity, $I_{1}+I_{2}=2$, occurs in the
subthreshold domain close to the saddle-node bifurcation (see diagonal line
d in Fig.3). These curves show that the relative phase control of the
subharmonics can be achieved via the control of the relative output
intensities, provided that an independent control of both cavity detunings
is implemented (see section V).

\begin{itemize}
\item  {\large Phase-locking}
\end{itemize}

The numerical phase $\eta $ related to the difference between the signal and
idler phases agrees extremely well with the mean-field prediction in Eqs.
(20b),( 22b) or (23),(24b) for any values of the detuning and cavity loss
parameter, but the numerical phase $\mu $ related to the sum of the signal
and idler phases is found to depart significantly from the mean-field value
(20a) when $\Delta _{1,2}/\kappa _{1,2}\succsim 1$. We have checked
numerically the self phase-locking effect of the signal-idler output, by
varing randomly the initial phase of the signal and idler noise. This
general result is illustrated in {\bf Fig.6}, for $\Delta _{1}/\kappa
_{1}=\Delta _{2}/\kappa _{2}$ where the phase difference $\eta $ follows
randomly the initial values (blank circles), when $\left| \chi \right| =0$,
as expected for a conventional DRO. However, yet for a vanishingly small
value $S=0.001$, the phase difference $\eta $ locks to $0$ or $\pi $ (mod $%
k\pi $) for $\xi =0$ and the sum phase locks to a constant value determined
only by the pumping rate.

\begin{itemize}
\item  {\large Time-dependent solutions}
\end{itemize}

For $\Delta _{1,2}\neq 0$ and $\Delta _{1}/\kappa _{1}=\Delta _{2}/\kappa
_{2,}$ the periodic solution remains generally stable on a large range of
the input intensity. For instance, for $\left| \chi \right| =0.2$ and $%
\overline{\Delta }=1$ (case of line c in Fig.2)$,$ the periodic solution is
stable until the pump parameter reaches $\alpha _{p}/\alpha _{th}\thicksim
12 $ , well above any experimentally achievable input. The period of the
amplitude oscillation in the vicinity of $I_{H}^{num}$ is found to be $T\sim
300$ $\tau $. But the case $\left| \Delta _{1}/\kappa _{1}-\Delta
_{2}/\kappa _{2}\right| \neq 0$ leads to a more complex dynamics, which
depends on the nonlinear coupling $\left| \chi \right| $ and departs from
the meanfield predictions .

In the zero-detuning case, no time-dependent solutions are found for
realistic pumping rate ($N<25$) for equal cavity loss $\kappa _{1}=\kappa
_{2}$. Differently, the case $\kappa _{1}\neq \kappa _{2}$ displays
time-dependent regimes, for $\left| \chi \right| \succsim 0.2$. In the case $%
\kappa _{1}/\kappa _{2}=2.5,\left| \chi \right| =0.2$ shown in {\bf Fig.7,}
a periodic regime arises at $I_{in}\cong 2$ $I_{th}$ and a weak chaotic
pulsating behaviour is found at $I_{in}\cong 4$ $I_{th}$, with a pulsating
signal intensity. Vanishingly small detunings ($\Delta _{1,2}$) however give
rise to steady state solutions satisfying relations (16).

The numerical solutions associated with $\Delta _{1,2}=0,$ $\Delta
_{2,1}\neq 0,$ are not stationary, as predicted by the meanfield model; they
are periodic with time for any input above threshold.

\begin{itemize}
\item  {\large Influence of SHG phase-mismatch }
\end{itemize}

We have studied the influence of a moderate SHG phase mismatch ($\xi \neq 0$%
) on the dynamical behaviour of the system. This phase mismatch is usually
controlled by the temperature or the angular orientation of the SHG crystal.
It provides a control of the strength of the nonlinear coupling parameter $%
\left| \chi \right| $ , via the relation (10), and offsets the value to
which the phase difference $\eta $ locks (Fig.6). For instance, a small
phase $\xi \leq 1$ does not significantly change the idler amplitudes of
Fig.3 but induces a slight imbalance of the signal and idler intensity ratio
compared to the ratio (19) and slightly reduces the self-locking range. The
relation $\mu =0$ (mod $\pi )$ is still valid and $\eta $ locks to $\xi $.
In some undesirable operating conditions, for instance, when the system
approaches a Hopf bifurcation for a given pumping rate and detuning set (see
Fig.2 for $S=0.2,\xi =0$), a small amount of phase mismatch produces a shift
of the periodic oscillation threshold towards higher pump rate.\
Differently, for $\stackrel{\_}{\Delta }=0$ and the same other parameters as
for curve (c) of Fig.2, a small phase mismatch ($\xi =0.1$) leads to a slow
periodic regime ($T\sim 4\times 10^{+6}$ $\tau $ at $\alpha _{in}/\alpha
_{th}=1.2$) which is suppressed when vanishingly small detunings (always
present in practical devices) are introduced.

From the simulations carried out, we conclude that a moderate SHG
phase-mismatch does not change significantly the main bifurcation dynamics
studied up to this point, except for the $\Delta _{1,2}=0$\ case.

Finally let us point out that the SPL-DRO solutions expanded up to the
second-order in Eq. (9) (ML2 approximation) agree very well with the exact
solutions in almost the whole range of the pump intensity and detunings
below the Hopf bifurcation threshold, except for the singular case $\Delta
_{1,2}=0$ corresponding to the transition from subcriticality to
supercriticality, where it fails to converge. Only the propagation model can
solve this case, which requires a double-precision computation due to the
slow convergence associated to a critical slowing-down phenomenon.

\section{\protect\bigskip {\bf SPL-TRO case}}

In the case of SPL-TRO, the meanfield equations (13b)-(13c) are still valid,
except that $I_{p}=\alpha _{p}^{2}$ denotes now the circulating intracavity
pump intensity. The intracavity pump amplitude obeys the boundary condition
(11a), leading to

\begin{equation}
\lbrack 1-r_{p}\exp (i\Delta _{p})]A_{p}-ir_{p}\exp (i\Delta
_{p})A_{1}A_{2}=A_{in}  \eqnum{25}
\end{equation}
where the pump detuning may be arbitrarily large ($\Delta _{p}\leq 2\pi )$ .
The relation between the input pump parameter $I_{in}=\left| A_{in}\right|
^{2}=\alpha _{in}^{2},$ and the cavity pump intensity is obtained, for an
arbitrary phase of the input field

\begin{equation}
I_{in}=(1+r_{p}^{2}-2r_{p}\cos \Delta _{p})\alpha _{p}^{2}+r_{p}^{2}\alpha
_{1}^{2}\alpha _{2}^{2}+2r_{p}\alpha _{1}\alpha _{2}\alpha _{p}[\sin (\Delta
_{p}-\mu )+r_{p}\sin \mu ]  \eqnum{26}
\end{equation}

where $\alpha _{1,}\alpha _{2}$ and $\mu $ can be deduced from Eqs.(16) and
(19-20). The threshold for oscillation is easily found from Eq. (26), on the
basis of simple considerations. Below the threshold, where $\alpha _{1,2}=0$%
, $I_{p}$ grows linearly as a function of $I_{in}$. When it reaches the
value given by Eq.(21), oscillation starts, which leads to the following
threshold input intensity

\begin{equation}
I_{th}^{SPL-TRO}=(1+r_{p}^{2}-2r_{p}\cos \Delta _{p})(I_{0}+\Delta
_{1}\Delta _{2})  \eqnum{27}
\end{equation}

In a practical TROs, although the pump finesse is lower than the finesse at
the subharmonic waves (typically an order of magnitude), the condition $%
\kappa _{p}\ll 1$ is still satisfied . If we consider a small enough pump
detuning and the specific case $\stackrel{\_}{\Delta }=\Delta _{1}/\kappa
_{1}=\Delta _{2}/\kappa _{2}$ , Eq.(27) reduces to the conventional TRO
threshold \cite{Debuisschert}

\begin{equation}
I_{th}^{TRO}=\left| A_{in}\right| _{th}^{2}=\kappa _{p}^{2}\kappa _{1}\kappa
_{2}(1+\stackrel{\_}{\Delta }^{2})(1+\stackrel{\_}{\Delta }_{p}^{2})
\eqnum{28}
\end{equation}
where $\stackrel{\_}{\Delta }_{p}=\Delta _{p}/\kappa _{p}$. {\bf Fig. 8}
shows the normalized intra-cavity pump and idler bifurcation diagrams as a
function of $\sqrt{\stackrel{\_}{I}_{in}}=\sqrt{I_{in}/I_{th}^{TRO}}$ when
condition (12a) holds and for $\Delta _{p}=0$. Only the details of the $%
\left| \chi \right| \alpha _{1}$ solutions in the vicinity of the threshold
are plotted in the top frame of Fig$.$ 8 that displays the subcriticality.\
The numerical and meanfield solutions of the idler amplitude as given by Eq.
(16) are confounded over a much larger input intensity range. The thick
solid and dashed lines for $\alpha _{p}$ correspond to the meanfield
solution (26), while the thin solid line is the exact numerical pump
solution. Notice that the numerical solution agrees very well with the
meanfield solution. As the input parameter is increased from threshold, $%
I_{p}$ decreases from the clamped value of the conventional TRO to a minimum
$I_{0}$ (Eq.17) for an input pump intensity $\stackrel{\_}{I}_{in}^{\min }$
given by

\begin{equation}
\stackrel{\_}{I}_{in}^{\min }=\left[ 1+(\kappa _{1}/\kappa _{p})(\stackrel{-%
}{\Delta }/\left| \chi \right| )^{2}\right] ^{2}/(1+\stackrel{-}{\Delta }%
^{2})  \eqnum{29}
\end{equation}
for $\Delta _{p}=0$. The meanfield model hence predicts that, unlike for
conventional TRO, the intracavity pump intensity is not clamped.\ For $%
\stackrel{-}{\Delta }\rightarrow 0$, $\stackrel{\_}{I}_{in}^{\min
}\rightarrow 1$, e.g. there is a transition to supercriticality as in the
SPL-DRO case.

In pump-detuned conventional TROs, subcriticality occurs only when the
condition $\stackrel{\_}{\Delta }_{p}\stackrel{\_}{\Delta }>1$ holds \cite
{Lugiato}, \cite{Debuisschert}. The experimental observation of this
subcriticality requires a high pumping level because of the pump detuning
dependence in Eq.(28) \cite{Ritchy} . It is interesting to study how the
subcriticality originating from the nonlinear OPO/SHG cascading would affect
the intrinsic detuned TRO bistability curve. In {\bf Fig.9} we have plotted
the numerically computed idler intensity versus the normalized input
intensity for $\left| \chi \right| =0,0.2,0.6$ and $\stackrel{\_}{\Delta }%
_{p}\stackrel{\_}{\Delta }=2$ . The subthreshold domain extends as $\left|
\chi \right| $ increases from zero, it has also increased, when compared to
the $\stackrel{\_}{\Delta }_{p}=0,$ $\left| \chi \right| =0.2$ case, shown
in Fig. 8. Furthermore, as $\left| \chi \right| $ increases, the numerical
threshold is smaller than the meanfield threshold given by Eq.(27).

We have also verified that the numerical amplitudes and phases agree with
the meanfield predictions for $\stackrel{\_}{I}_{in}<10$ and for detuning
values such that $\Delta _{j}/\kappa _{j}\precsim 1$. Far above threshold
and for larger detunings, only the phase difference relations remain exact,
like in the SPL-DRO case.

Furthermore, the detuning range $\stackrel{\_}{\Delta }$ for self-locking is
less dependent on the coupling strength than in the SPL-DRO case.

The high-pump-finesse SPL-TRO device does not exhibit the time-dependent
periodical solutions observed above threshold with the SPL-DRO as long as $%
\left| \chi \right| \leq 0.52$ (which corresponds roughly to equal OPO and
SHG crystal length for a PPLN) and for realistic maximum pumping rate $%
\stackrel{-}{I}_{in}=I_{in}/I_{th}^{SPL-TRO}$ \ $\leq 2500$, corresponding
to a Watt-level pump power and typical TRO pump thresholds in the mW range.
For larger coupling $\left| \chi \right| =0.6,$ periodic oscillations occur,
with a period varying in a very complex way when the input pump is
increased. Nevertheless no chaotic regime is observed in the range of
considered input pump.

In many practical experiments, one cannot avoid a weak pump resonance in
DROs due to the multiwavelength coatings of the mirrors. It is thus
interesting to investigate how a moderate pump resonance would affect the ($%
\Delta _{1},\Delta _{2}$) self-locking domain of Figs.5. The input\ external
pump intensity is kept equal to $I_{in}^{ext}=4\kappa _{1}\kappa _{2}$ to
provide a comparison with Figs.5 ( $\left| \chi \right| $ $=0.2$). Such a
pumping level corresponds to an internal TRO pumping rate $\stackrel{-}{I}%
_{in}/I_{th}^{TRO}=$ $8/\kappa _{p}=160$. In Fig. 10, the pump reflectivity
is $r_{p}=0.8$ with $\Delta _{p}=0$. Apart from the slight enlargement of
the self-locking domains, as compared to Fig.5, which is mainly attributed
to the high pumping rate, the intensities are smaller than for the DRO, also
the intensity distributions are strongly modified, with off-diagonal maxima.

In summary, the only improvement due to a moderate pump resonance is the
stabilization effect respective to the onset of temporal dynamics. The
extended self-phase locking range is paid back with lower output intensities.

\section{\bf Practical implementation of SPL-OPOs}

In order to avoid spurious cavity loss, the use of a dual-grating
quasi-phase matched periodically poled crystal is particularly well suited
to the implementation of SPL-OPOs. However, the grating period should be
accurately designed so as to phase-matched simultaneously both interactions
for the same chip temperature, even though the theoretical analysis predicts
a minor influence of SHG phase mismatch.

From an experimental point of view, it is desirable to implement 3:1
SPL-OPOs using the least constraining OPO configuration. Diagnosis methods
to check the high phase coherence between the subharmonics under SPL have to
be implemented. In the frequency domain, this can be achieved by monitoring
the beatnote between the signal wave and the externally frequency-doubled
idler wave (or between the summed subharmonics and the pump). Because under
SPL operation these two waves are frequency degenerate, the output signal
wave must be preliminary frequency-shifted by a suitable RF frequency $%
\omega _{RF}$ using, e.g., an acousto-optic modulator. When the OPO operates
within the locking range, the beatnote frequency should be fixed to $\omega
_{RF}$ and its power spectral density should approach a Dirac function.
Another equivalent method to check the phase coherence is to perform an
interferometric fringe pattern measurement by overlapping the two beams on a
slow detector\cite{Mason}. In the following, other indirect methods, based
on the theoretical analysis will be outlined.

As a starting point of the analysis, we have assumed perfect 3:2:1 frequency
ratios for the pump, signal and idler waves. In practical devices, such a
situation will be unlikely met at once. The major difficulty will come from
the fine (continuous) tuning of the signal and idler frequencies close
enough to the 2:1 degeneracy, in order to fall within the capture range of
the self-phase locking. For a fixed pump frequency, such a fine tuning is
usually performed via the temperature or angle tuning of the phase-matching.
Singly resonant devices (PRSROs) offer an easy and relaxed mode-hop free
frequency tuning because only one subhamonic wave is resonant, especially in
a dual-arm cavity configuration to control independently the pump and signal
resonances \cite{Dunn}. However from our analysis of the SPL-PRSRO with a
resonant signal wave, which also displays subcriticality, extremely small
self-locking ranges are predicted due to the weak coupling between the
injecting frequency-doubled idler and the resonant signal. While in a
conventional PRSRO the signal field is constrained to oscillate with a nil
cavity detuning, the nonlinear coupling is found to allow oscillation over a
small detuning range not exceeding the cavity linewidth. The increase of the
locking range versus the coupling parameter (e.g. the SHG crystal length) is
only moderate, even with $\left| \chi \right| =1$ (which would correspond to
a SHG crystal $\thicksim $ 1.5 times longer than the OPO crystal supposing
that both OPO and SHG interactions have the same nonlinearity magnitude) .
Due to this limited capture range, experimentally confirmed in ref \cite
{Boller}, 3:1 SPL-PRSROs \ would probably require an additional electronic
servo on the cavity length to operate as a stable divider device. A
convenient criterion for the assessment of SPL in PRSROs would be the slight
enhancement of the signal intensity when the nonlinear coupling is switched
on, compared to the slight decrease predicted for the SPL-DRO (Fig.3 ). A
possible way to realize such a switching is to have an OPO-only grating
section patterned beside the OPO/SHG dual-grating section on the
periodically-poled wafer.

From our theoretical study, the widest SPL range (a few cavity linewidths)
is obtained with the doubly resonant configuration (SPL-DRO) due to the
strong self-injection regime, with eventually a weak pump resonance
(SPL-TRO) to stabilize the device. Even though the amount of nonlinear
coupling required can be extremely small ( $\left| \chi \right| =0.01$ ) -
and such low level doubled idler can even be spontaneously generated via
non-phase matched or higher-order quasi-phase matching in PP single-grating
nonlinear OPO crystals - a coupling strength corresponding to $\left| \chi
\right| =0.1-0.2$ will ensure a robust self-phase locking of the subharmonic
wave. Preliminary single mode-pair operation of DRO/TROs usually require an
intensity sidelock servo to control the stability of the cavity length. This
sidelock servo compares the output signal (or idler) intensity to a stable
electronic voltage reference which sets the operating (usually non-zero)
signal and idler cavity detuning values. When a linear cavity is used, the
signal and idler (plus eventually the pump) detunings cannot be
independently controlled via the cavity length. It is then probable that the
oscillating mode pair will have equal normalized detunings (condition (12a))
that satisfy the minimum threshold.\ Under sidelock servo the transition
from conventional to SPL states is accompanied necessarily by a detuning
transition (see Fig.3). It is then important to set the sidelock reference
voltage as close as to the maximum fringe intensity in order that the new
detuning value does not exceed the allowed subthreshold range. In the case
of a well resolved DRO mode pair cluster, the observation of these
subthreshold states, and the associated broadened mode pair fringe, should
be made possible via adiabatic cavity length tuning. Such an observation
would be an indirect diagnosis of SPL. But it is necessary to have an
independent control of the signal and idler detunings to explore the full
allowed range of ($\Delta _{1},\Delta _{2}$) detunings depicted in Figs.5
and 10. A dual-cavity DRO/TRO design would then be appropriate. The control
of these detunings allows the control of the relative phase between the
pump, signal and idler. The output subharmonic intensity ratio $I_{1}/I_{2}$
(see Eq.19) can be used as an error signal for the relative phase control.

\section{\bf Conclusions and outlook}

We have theoretically demonstrated that resonant $\chi ^{(2)}:\chi ^{(2)}$
nonlinear cascaded OPO/SHG processes induce a self injection-locking between
the subharmonic waves of an OPO leading to the self-phase locking of the
three interacting waves, unlike in a conventional OPO for which the absolute
phases of the signal and idler are undetermined. The theoretical treatment
encompasses the meanfield model as well as a full propagation model. The
doubly and triply resonant oscillator configurations lead to the widest self
phase-locking ranges. The main conclusions are: a) the minimum threshold of
oscillation of these devices are identical to the conventional devices; b)
the nonlinear cascading leads to a subcritical behaviour, even in the case
of a DRO or a PR-SRO, and can lead to the occurrence of self-pulsing
instabilitities. This subcriticality is different from the standard
subcriticality reported in pump detuned conventional TROs. The nonlinear
coupling removes the detuning constraints of conventional systems, allowing
for a potentially accurate control of the relative phase between the
subharmonic waves. The range of allowable detunings over which the field
phases are locked depends on the magnitude of the nonlinear coupling and on
the self-injection regime. While this range is smaller than the cavity
linewidth for PRSROs (weak injection regime), it spans over several cavity
linewidths under strong injection-locking regime obtained in signal/idler
resonant devices (DRO/TROs). The SPL-DRO/TRO give rise to a richer dynamics
than the singly resonant PRSRO for which no occurence of a Hopf bifurcation
is found in the whole range of the system parameters. It is thus interesting
to extend the theoretical model by including diffraction effects in order to
investigate the possible occurrence of new spatio-temporal dynamics, \cite
{Meanf3},\cite{Oppo}. These self-phase locked OPOs will be useful tools for
applications requiring a high degree of optical phase coherence between
optical harmonic waves, such as precision optical measurements in the mid-IR
or Fourier synthesis of ultra-short optical pulses.

The model developed can be easily extended to the study of divide-by-4
SPL-OPOs based on the cascading OPO/OPO processes $4\omega \rightleftarrows
2\omega \rightleftarrows \omega $ which has the potential to generate up to
8 phase-locked harmonic waves by additional up-conversion processes. Such a
strong nonlinearly coupled system can be viewed as a secondary degenerate
OPO (DOPO) embedded in a primary DOPO. Our future work will be directed to
the theoretical investigation of the stability of 4:1 OPO dividers. A
classical signature of the system, derived from the meanfield analysis, is
the clamping of the secondary pump $2\omega $ to the threshold power for the
fundamental oscillation.

The present study of 3:1 OPO dividers, which makes the simplest assumption
of exact 3:2:1 frequency ratios, arouses another interrogation. An
interesting situation not considered regards the behaviour of the nearly 3:1
OPO/SHG system when the frequency ratios slightly departs from the perfect
3:2:1 division ratio by a radio-frequency quantity $\delta <<\omega $, e.g.
when $3\omega \rightarrow 2\omega -\delta ,\omega +\delta $. The frequency
difference beween the signal wave and the doubled idler is then $\left|
2\omega _{i}-\omega _{s}\right| =3\delta $. When the doubled idler frequency
does not match one of the cavity eigenmode frequencies, the operation of the
OPO would be merely that of a conventional DRO. But if the OPO is tuned such
that $3\delta =FSR_{s}\thickapprox c/\Lambda ,$ $FSR_{s}$ being the signal
free spectral range of the cavity, then the doubled idler will be enhanced
to a point where it may lead to the creation of a new mode pair with
frequencies ($2\omega +2\delta ,\omega -2\delta $) and so on. The parametric
gain bandwidth of OPOs extending usually over several THz or several tens of
THz (in case of a wavelength non-critical phase matching), a multitude of
self-phase locked mode pairs equally spaced by $3\delta $ may potentially
oscillate, provided that the pumping level is sufficiently high. Such a
complex system opens the prospect of building a mode-locked, dual-band OPO
frequency comb generator using cascaded second-order nonlinearities as the
passive mode-locking mechanism. In the time domain, the output of such a
system would consist of a train of short optical pulses with a repetion rate
set by the $FSR_{s}$ intermode spacing, provided that the relative phase
between adjacent mode pairs is preserved and group velocity dispersion is
compensated. We note that a similar cw-DRO running near frequency degeneracy
with thousands of mode pairs actively locked by an intracavity electro-optic
phase modulator has been recently reported, with a striking passive output
stability feature compared to a conventional quasi-degenerate single-mode
pair DRO\cite{Hall}.\bigskip

\begin{acknowledgement}
This work is partially supported by an INCO-Copernicus European network
program (contract n$%
{{}^\circ}%
$ ERBIC 15CT 98 0814 ).
\end{acknowledgement}

\section{\protect\appendix}

\section{Linear stability analysis of the SPL-DRO}

With $A_{j}(t)\equiv $ $A_{j}(t,0),$ Eqs. (9-11) give rise to the mapping
equations

\begin{eqnarray}
A_{2}(t+\tau ) &=&r_{2}e^{i\Delta _{2}}[A_{2}(t)+iA_{p}(t)A_{2}^{\ast
}(t)+i\chi ^{\ast }A_{1}^{2}(t)]]  \eqnum{A1} \\
A_{1}(t+\tau ) &=&r_{1}e^{i\Delta _{2}}[A_{1}(t)+iA_{p}(t)A_{1}^{\ast
}(t)+i\chi A_{2}(t)A_{1}^{\ast }(t)]  \eqnum{A2} \\
A_{p}(t) &=&A_{in}  \eqnum{A3}
\end{eqnarray}

the stationary solutions of which are deduced for large $r_{1,2}$ and small $%
\Delta _{1,2}.$ (See Eqs.(19),(22)-(23)).

The linear stability analysis consists in assuming small deviations $\delta
A_{1,2}(t)$ from the stationary solutions $\overline{A}_{1,2}$ $,$

\begin{equation}
A_{1,2}(t)=\overline{A}_{1,2}+\delta A_{1,2}(t)  \eqnum{A4a}
\end{equation}

\begin{equation}
\delta A_{1,2}(t)=%
\mathrel{\mathop{\sum }\limits_{\lambda }}%
\delta A_{1,2}(\lambda )e^{\lambda t}  \eqnum{A4b}
\end{equation}

where $\lambda $ may be complex. The stationary solutions $\overline{A}
_{1,2} $ are stable only if the real part of any $\lambda $ is negative. At
the instability threshold, therefore the system may undergo a Hopf
bifurcation ( $\lambda _{0}=\pm i\beta ),$ so that the intensitiy oscillates
with time at angular frequency $\beta .$

Eqs. (A1-A4) lead to a linearized system of four equations, the determinant
of which satisfies

\begin{equation}
D=\left|
\begin{array}{llll}
\Lambda +\alpha _{1} & -i\chi _{1}\overline{A}_{2} & -i\chi _{1}\overline{A}
_{1}^{*} & -i\chi _{1}^{{}}A_{in,1} \\
i\chi _{1}^{*}\overline{A_{2}^{*}} & \Lambda +\alpha _{1}^{*} &
iA_{_{in,1}}^{*} & i\chi _{1}^{*}\overline{A}_{1}^{{}} \\
-2i\chi _{2}^{*}\overline{A}_{1} & -iA_{in,2} & \Lambda +\alpha _{2} & 0 \\
iA_{in,2}^{*} & 2i\chi _{2}^{{}}\overline{A}_{1} & 0 & \Lambda +\alpha
_{2}^{*}
\end{array}
\right| =0  \eqnum{A5}
\end{equation}

with the notations
\begin{equation}
\Lambda =e^{\lambda \tau }-1  \eqnum{A6}
\end{equation}

and
\begin{eqnarray}
\alpha _{1,2} &=&\kappa _{1,2}-i\Delta _{1,2}  \eqnum{A7a} \\
\chi _{1} &=&e^{i\Delta _{1}}\chi r_{1},\chi _{2}=e^{-i\Delta _{1}}\chi r_{2}
\eqnum{A7b} \\
A_{in,1,2} &=&r_{1,2}e^{i\Delta _{1,2}}A_{in}  \eqnum{A7c}
\end{eqnarray}

Then, the eigenvalues $\Lambda $ are solutions of the quartic characteristic
equation
\begin{equation}
\Lambda ^{4}+\Phi _{3}\Lambda ^{3}+\Phi _{2}\Lambda ^{2}+\Phi _{1}\Lambda
+\Phi _{0}=0  \eqnum{A8}
\end{equation}

where all the coefficients $\Phi _{i}$ are real

\begin{eqnarray}
\Phi _{3} &=&-2(\kappa _{1}+\kappa _{2})  \eqnum{A9a} \\
\Phi _{2} &=&\left| \alpha _{1}\right| ^{2}+\left| \alpha _{2}\right|
^{2}+4\kappa _{1}\kappa _{2}-2r_{1}r_{2}\left| \overline{A}_{in}\right|
^{2}\cos (\Delta _{1}-\Delta _{2})-\left| \chi r_{2}\overline{A}_{2}\right|
^{2}  \nonumber \\
&&+4r_{1}r_{2}\left| \chi \overline{A}_{1}\right| ^{2}\cos (\Delta
_{1}+\Delta _{2})  \eqnum{A9b} \\
\Phi _{1} &=&2r_{1}r_{2}\left| \overline{A}_{in}\right| ^{2}\left[ (\Delta
_{1}+\Delta _{2})\sin (\Delta _{1}+\Delta _{2})-(\kappa _{1+}\kappa
_{2})\cos (\Delta _{1}-\Delta _{2})\right] +2(\kappa _{2}\left| \alpha
_{1}\right| ^{2}+\kappa _{1}\left| \alpha _{2}\right| ^{2})  \nonumber \\
&&-2\kappa _{2}\left| \chi r_{2}\overline{A}_{2}\right| ^{2}+4\left| \chi
r_{1}r_{2}\overline{A}_{1}\right| ^{2}\left[ (\kappa _{1}+\kappa _{2})\cos
(\Delta _{1}+\Delta _{2})+(\Delta _{1+}\Delta _{2})\sin (\Delta _{1}+\Delta
_{2})\right]  \nonumber \\
&&-ir_{1}r_{2}\left| \chi \right| ^{2}\cos \Delta _{2}(\overline{A}_{in}%
\overline{A}_{2}^{\ast }\overline{A}_{1}^{\ast }-c.c.)+3r_{1}r_{2}\left|
\chi \right| ^{2}\sin \Delta _{2}(\overline{A}_{in}\overline{A}_{2}^{\ast }%
\overline{A}_{1}^{\ast }+c.c.)  \eqnum{A9c} \\
\Phi _{0} &=&\left| \alpha _{1}\right| ^{2}\left| \alpha _{2}\right|
^{2}-2r_{1}r_{2}\left| \overline{A}_{in}\right| ^{2}\left[ (\kappa
_{1}\kappa _{2}+\Delta _{1}\Delta _{2})\cos (\Delta _{1}-\Delta
_{2})+(\Delta _{1}\kappa _{2}-\Delta _{2}\kappa _{1})\sin (\Delta
_{1}-\Delta _{2})\right]  \nonumber \\
&&+4r_{1}r_{2}\left| \chi \overline{A}_{1}\right| ^{2}\left[ (\kappa
_{1}\kappa _{2}-\Delta _{1}\Delta _{2})\cos (\Delta _{1}+\Delta
_{2})+(\Delta _{1}\kappa _{2}+\Delta _{2}\kappa _{1})\sin (\Delta
_{1}+\Delta _{2})\right]  \nonumber \\
&&-\left| \chi r_{1}\alpha _{2}\overline{A}_{2}\right| ^{2}+\left( 4\left|
\chi r_{2}\overline{A}_{1}\right| ^{2}-r_{2}^{2}\left| \overline{A}%
_{in}\right| ^{2}\right) \left( \left| \chi r_{1}\overline{A}_{1}\right|
^{2}-r_{1}^{2}\left| \overline{A}_{in}\right| ^{2}\right)  \eqnum{A9d}
\end{eqnarray}
.

The roots for $\Lambda $ and consequently the eigenvalues $\lambda $ have
been calculated for various detunings $\Delta _{1,2}$ as a function of the
input amplitude $A_{in}$ for the trivial solutions $\overline{A}_{1,2}=0$
and the non trivial solutions ($\overline{A}_{1,2})_{\pm }$ deduced from
Eqs.(16) and (20).

The trivial solution is verified to be stable for any input pump intensity
below $I_{th}$(Eq.18).

On the other hand the solution $X_{_{-}}$ is found to be unstable with
respect to any constant perturbation for any detuning and input pump
intensity: Indeed, in this case, Eq.( A8) has a root $\Lambda =0,$i.e, $%
\lambda =0,$ because the constant coefficient $\Phi _{0}$ is identically
nil, when replacing $X_{_{-}}$ by its expression (16).

Differently, the upper branch displays two sets of complex conjugate
eigenvalues, ($\sigma _{1,2}\pm i\beta _{1,2})$ with negative real parts for
a range of the input pump intensity, lying from $I_{0}$ to $I_{H}$ where the
system undergoes a Hopf bifurcation ($\sigma _{1}=0)$. For $I_{p}>$ $I_{H},$
the signal and idler intensities oscillate with time at angular frequency $%
\beta _{1}$. Therefore the upper branch is stable for $I_{0}$%
\mbox{$<$}%
$I_{p}<I_{H}.$

The Hopf bifurcation threshold intensity $I_{H}$ may be either smaller or
larger than the threshold intensity $I_{th}$, depending on detunings $\Delta
_{1}$ and $\Delta _{2}$. Let us introduce the parameter $\mu _{H}$ that
measures the departure between the threshold for lasing $I_{th}$ and the
Hopf bifurcation threshold as

\begin{equation}
\mu _{H}=(I_{th}-I_{H})/I_{th}  \eqnum{A10}
\end{equation}

The variation of $\mu _{H}$ is presented in {\bf Fig. 11} as a function $%
\Delta _{1}=\Delta $ either with $\Delta _{1}=\Delta _{2}$ in curve (a) or $%
\Delta _{1}=\frac{1}{2}\Delta _{2}$ in curve (b).

For $\Delta _{1}=\Delta _{2},$ the Hofp bifurcation ($\sigma _{1}=0,\sigma
_{2}<0)$ occurs above threshold only for $\Delta \lesssim \kappa .$
Therefore lasing stationary solutions can be reached from rest only for
detunings smaller than the cavity loss coefficient. Otherwise, lasing at
larger detunings can be reached, when increasing adiabatically the detunings
from values smaller than $\kappa $. In the other case, $\Delta _{1}=\frac{1}{%
2}\Delta _{2}$, $\mu _{H}$ is positive for any $\Delta _{1}\lesssim 5\kappa $%
. For higher detuning, $\sigma _{2}$ reaches zero at an input intensity much
smaller than the value at which $\sigma _{1}=0$ crosses zero. This causes an
abrupt decrease of $I_{H},$ so that $\mu _{H}$ suddenly becomes negative for
$\Delta _{1}\geqslant 5\kappa ,$ as shown in the curve (b) of {\bf Fig. 11}.

Finally, the angular frequency $\beta _{H}$ at the Hopf bifurcation
threshold, not reported here, is found to vary proportionally to the
detuning.

\newpage

\begin{center}
{\LARGE Figure captions}
\end{center}

\bigskip \noindent

\noindent {\bf Fig.1: }Schematic ring cavity model of SPL-OPOs. All
intracavity losses are lumped into the output mirror transmissivities $t_{j}$%
.

\noindent {\bf Fig.2:} Comparison of the meanfield and numerical stationary
solutions of SPL-DRO, with $\kappa _{1,2}=\Delta _{1,2}=0.005$, as a
function of the scaled pump input amplitude ($\alpha _{th}=\sqrt{I_{th}}$).
(a): meanfield upper branch (The dotted portion corresponds to unstable
solutions), (a'): meanfield lower branch . (b): numerical solution for $%
\left| \chi \right| =0.05$. (c): numerical solutions for $\left| \chi
\right| =0.2$.

\noindent {\bf Fig.3:}Numerical SPL-DRO self locking ranges as a function of
the detuning $\Delta _{1}/\kappa _{1}(=\Delta _{2}/\kappa _{2})$ , for $%
I_{in}=4\kappa _{1}\kappa _{2}$, with $\left| \chi \right| =0.01$ in (a), $%
0.05$ in (b), $0.1$ in (c), $0.2$ in (d) and $0.25$ in (e). The dashed line
is for $\left| \chi \right| =0$. The parameters are $\kappa _{1,2}=0.005$.

\noindent {\bf Fig.4:} Time response of the SPL-DRO stationary state
operating at $\left\{ \left| \chi \right| =0.2,\stackrel{\_}{\Delta }%
_{opt}=4\right\} $, under step-wise linear detuning jumps $\stackrel{\_}{%
\Delta }_{opt}(1\pm \beta )$, with modulation index $\beta =0.109$ (curve
a). Curve (b) gives the idler amplitude response and curve (c) the phase
difference response. The re-capture range of the detuning perturbation is $%
\beta _{\max }=0.12$.

\noindent {\bf Fig.5:} Contour plots of the stationary SPL-DRO signal (a),
idler (b), and total (c) intensities and phase difference $\eta $ (d) in the
2D detunig plane, obtained from the numerical computation with $\kappa
_{1,2}=0.005$, $\left| \chi \right| =S=0.2$, $I_{in}=4\kappa _{1}\kappa _{2}$
. The intensities are scaled to $I_{1,2}=2\kappa _{2,1}$. The pale blue
domain are trivial solutions.

\noindent {\bf Fig.6:} Numerically computed distribution of stationary
signal/idler phase difference as defined in eq.(14d) versus the initial
random phase for $\Delta _{1,2}=\kappa _{1,2}=0.005$, $\alpha _{in}/\alpha
_{th}=2$ and $S=0.001$, for a SHG phase mismatch $\xi =\pi $ (i.e $\chi =0$)
in blank circles; $\xi =0$ in solid black circles and $\xi =3\pi /2$ in
black triangles. Note that each phase data point is associated to the same
stationary signal-idler intensities and the same phase sum $\mu $.

\noindent {\bf Fig.7:} Time-dependent numerical solutions for $\kappa
_{1}=0.005$, $\kappa _{2}=0.002$, $\Delta _{1,2}=0$, $\chi =0.2$, $%
I_{in}=1.96\kappa _{1}\kappa _{2}$ (bottom frame) and $I_{in}=4\kappa
_{1}\kappa _{2}$ (top frame).

\noindent {\bf Figs.8:} Meanfield and numerical SPL-TRO stationary solutions
for $\Delta _{p}=0$, $r_{p}=0.9$ and the same other parameters as for Fig.2.%
{\bf \ }The top frame shows{\bf \ }details of the hysteresis loop of the
idler amplitude in the vicinity of the input pump threshold. The thin lines
correspond to the numerical solutions, interrupted by the vertical thin
dashed line. The thick lines correspond to the meanfield solutions. The
thick dashed line in the bottom frame is the meanfield unstable branch of $%
\alpha _{p}/\alpha _{th}$ ($\alpha _{th}=\sqrt{\kappa _{1}\kappa _{2}+\Delta
_{1}\Delta _{2}}$), and the thinner horizontal dashed line shows the
conventional TRO pump clamping.

\noindent {\bf Fig.9:} Subcriticality in the strongly pump-detuned
conventional TRO ($S=0$) and SPL-TRO ($S=0.2,0.6$) for $r_{p}=0.9,$ $%
\stackrel{\_}{\Delta }=1,\stackrel{\_}{\Delta }_{p}=2$ and same other
parameters as in Fig.2. The thresholds located at the vertical arrows, are
lower than the one given by eq.(27).

\noindent {\bf Fig.10: }Contour plots of the numerical stationary SPL-TRO
signal (a), idler (b), and total (c) intensities and phase difference $\eta $
(d) in the 2D detuning plane with $r_{p}=0.8$, $\Delta _{p}=0$, $\kappa
_{1,2}=0.005$, $\chi =S=0.2$, $I_{in}=4\kappa _{1}\kappa _{2}$ . The
intensities are scaled to $I_{1,2}=2\kappa _{2,1}$. In the pale blue domain
there is no lasing. For $r_{p}=0.9$, the self locking domain extends over
the whole 2D frame.

\noindent {\bf Fig.11: }Plots of the reduced Hopf bifurcation thresholds $%
\mu _{H}=(I_{th}-I_{H})/I_{th}$ versus the normalised idler detuning $\Delta
_{1}/\kappa _{1}$, for $\Delta _{2}=\Delta _{1}$ (solid line (a)); $\Delta
_{2}=2\Delta _{1}$ (solid line (b)). The cavity loss parameters used are $%
\kappa _{1,2}=\kappa =0.005$ and $S=0.2$.


\begin{references}
\bibitem{Hansch/Telle}  H.R.\ Telle, D.\ Meschede, and T.W.\ H\"{a}nsch,
Opt.\ Lett. {\bf 15}, 532 (1990).

\bibitem{Wong}  N.C.\ Wong, Opt.\ Lett. {\bf 17}, 1155 (1992).

\bibitem{Hall/Hansch}  S.A.\ Diddams {\it et al}, Phys.\ Rev. Lett. {\bf 84}%
, 5102 (2000).

\bibitem{HanschOC}  T.W.\ H\"{a}nsch, Opt.\ Commun. {\bf 80}, 71 (1990).

\bibitem{Mukai}  T.\ Mukai, R.\ Wynands and T.W.\ H\"{a}nsch, Opt.\ Commun.
{\bf 95}, 71 (1993).

\bibitem{Kobay1}  Y.\ Kobayashi and K. Torizuka, Opt.\ Lett. {\bf 25}, 856
(2000).

\bibitem{Siegman}  A.E.\ Siegman, in {\it Lasers} (see chapt. {\bf 27,29}),
Univ. Science Books ed.(Sausalito, California ,1986).

\bibitem{Hyodo}  M.\ Hyodo, N.\ Onodera, and K.S.\ Abedin, Opt.\ Lett. {\bf %
24}, 303 (1999).

\bibitem{Pfister}  O.\ Pfister {\it et al}, Opt.\ Lett. {\bf 21}, 1387
(1997).

\bibitem{Touahri}  D.\ Touahri {\it et al}, Opt.\ Commun. {\bf 133}, 471
(1997).

\bibitem{Nee1}  P.T.\ Nee and N.C.\ Wong, Opt.\ Lett. {\bf 23}, 46 (1998).

\bibitem{Bernard}  J.E.\ Bernard, B.G.\ Whitford, and L.\ Marmet, Opt.\
Lett. {\bf 24}, 98 (1999).

\bibitem{Graham}  R.\ Graham and H.\ Haken, Zeit. f\"{u}r Phys. {\bf 210},
276 (1968).

\bibitem{Lee}  D.\ Lee and N.C.\ Wong, Opt. Lett. {\bf 17}, 13 (1992).

\bibitem{Slyusarev}  S.\ Slyusarev, T.\ Ikegami and S.\ Oshima, Opt.\ Lett.
{\bf 24}, 1856 (1999).

\bibitem{Douillet}  A.\ Douillet et al, to appear in IEEE Trans.\ Meas. \&
Instr. (special volume on CPEM conf., Sydney, 2000).

\bibitem{Mason}  E.\ J.\ Mason and N.C.\ Wong, Opt.\ Lett. {\bf 23}, 1733
(1998).

\bibitem{Boller}  D.-H.\ Lee {\it et al}, Opt.\ Express {\bf 5}, 114 (1999).

\bibitem{Nabors}  C.D.\ Nabors et al, Opt.\ Lett. {\bf 14}, 1189 (1989).

\bibitem{NaborsOPO}  C.D.\ Nabors, S.T.\ Yang, T.\ Day, and R.L.\ Byers, J.\
Opt.\ Soc.\ Am.\ {\bf B7}, 815 (1990).

\bibitem{Fabre-Wong}  C.\ Fabre, E.J.\ Mason, and N.C.\ Wong, Opt. Commun.
{\bf 170}, 299 (1999).

\bibitem{Tang}  P.P.\ Bey and C.L.\ Tang, IEEE J.\ Quantum Electron. QE-8,
361 (1972).

\bibitem{Cheung}  G.T.\ Moore, K.\ Koch, and E.C.\ Cheung, Opt.\ Commun.
{\bf 113}, 463 (1995); J.\ Opt.\ Soc.\ Am.\ {\bf B12}, 2268 (1995).

\bibitem{Moore}  G.T.\ Moore and K.\ Koch, IEEE J.\ Quantum Electron.\ {\bf %
29}, 961 (1993).

\bibitem{Dikmelik}  Y. Dikmelik and O. Ayt\"{u}r, IEEE J. Quantum Electron.
{\bf 35}, 897 (1999).

\bibitem{Schiller-casc}  A.G.\ White et al, Phys.\ Rev.\ {\bf A55}, 4511
(1997).

\bibitem{Marte}  M.A.M.\ Marte, J.\ Opt.\ Soc.\ Am.\ {\bf B12}, 2296 (1995).

\bibitem{Kobayashi}  L.\ Zhang, Q. Zheng, and T. Kobayashi, J.\ Opt.\ Soc.\
Am.\ {\bf B14}, 979 (1997).

\bibitem{Lugiato}  L.A.\ Lugiato, C.\ Oldano, C.\ Fabre, E.\ Giacobino and
R.J.\ Horowicz, Il Nuovo Cimento {\bf 10D}, 959 (1988).

\bibitem{Debuisschert}  T.\ Debuisschert, A. Sizmann, E.\ Giacobino, and C.\
Fabre, J.\ Opt.\ Soc.\ Am.\ {\bf B10}, 1668 (1993).

\bibitem{Meanf1}  L.\ Lugiato, in {\it Progress in Optics}, .E.\ Wolf
ed.(North-Holland, Amsterdam, 1984), Vol.\ XXI, p.69.

\bibitem{Meanf2}  M. Le Berre, A.S. Patrascu, E Ressayre, A. Tallet and
N.I.\ Zheleznykh, Chaos, Solitons, \& Fractals {\bf 4}, 1389 (1994).

\bibitem{Meanf3}  M.\ Tlidi, M. Le Berre, E.\ Ressayre, A.\ Tallet, and L.
Di Menza, Phys.\ Rev.\ {\bf A61}, 043806 (2000).

\bibitem{OFCG}  L.R.\ Brothers, D.\ Lee, and N.C.\ Wong, Opt.\ Lett.\ {\bf 19%
}, 245 (1994).

\bibitem{Ritchy}  C.\ Ritchy, K.I. Petsas, E.\ Giacobino, C.\ Fabre and L.\
Lugiato, J.\ Opt.\ Soc.\ Am.\ {\bf B12}, 456 (1995).

\bibitem{Dunn}  F.G.\ Colville, M.J.\ Padgett, and M.H.\ Dunn, Appl.\ Phys.\
Lett. {\bf 64}, 1490 (1994).

\bibitem{Oppo}  G.-L.\ Oppo, M.\ Brambilla, and L.A.\ Lugiato, Phys.\ Rev.\
{\bf A49}, 2028 (1994).

\bibitem{Hall}  S.A.\ Diddams, L.-S.\ Ma, J.\ Ye, and J.L.\ Hall, Opt. Lett.
{\bf 24}, 1747 (1999).
\end{references}
\end{document}